\documentclass[letterpaper]{article}
\usepackage{aaai23}
\usepackage{times}
\usepackage{helvet}
\usepackage{courier}
\usepackage[hyphens]{url} 
\usepackage{graphicx}
\def\UrlFont{\rm} 
\usepackage{natbib}  
\usepackage{caption}  
\usepackage{cite}
\usepackage{amsmath,amssymb,amsfonts}
\usepackage{algorithm,algorithmic}

\usepackage{subfigure}
\usepackage{textcomp}
\usepackage[dvipsnames]{xcolor}
\usepackage{filecontents}
\usepackage{url}
\usepackage{colortbl}
\usepackage{makecell}
\usepackage{pgfplots}
\usepackage{multirow}
\usepackage{caption,latexsym}
\usepackage{wrapfig}
\usepackage{tikz}
\usepackage{slashbox}

\newcommand{\eat}[1]{}
\newtheorem{definition}{Definition}

\begin{document}
\title{Mastery Guided Non-parametric Clustering to Scale-up Strategy Prediction}
\author{
Anup Shakya\\
ashakya@memphis.edu\\
University of Memphis
}
\author {
    Anup Shakya, Vasile Rus, Deepak Venugopal\\
}
\affiliations {
    University of Memphis\\
    \{ashakya, vrus, dvngopal\}@memphis.edu
}

\maketitle

\begin{abstract}
Predicting the strategy (sequence of concepts) that a student is likely to use in problem-solving helps Adaptive Instructional Systems (AISs) better adapt themselves to different types of learners based on their learning abilities. This can lead to a more dynamic, engaging, and personalized experience for students. To scale up training a prediction model (such as LSTMs) over large-scale education datasets, we develop a non-parametric approach to cluster symmetric instances in the data. Specifically, we learn a representation based on {\em Node2Vec} that encodes symmetries over {\em mastery} or skill level since, to solve a problem, it is natural that a student's strategy is likely to involve concepts in which they have gained mastery. Using this representation, we use DP-Means to group symmetric instances through a coarse-to-fine refinement of the clusters. We apply our model to learn strategies for Math learning from large-scale datasets from MATHia, a leading AIS for middle-school math learning. Our results illustrate that our approach can consistently achieve high accuracy using a small sample that is representative of the full dataset. Further, we show that this approach helps us learn strategies with high accuracy for students at different skill levels, i.e., leveraging symmetries improves fairness in the prediction model.
\end{abstract}
\section{Introduction}

\begin{figure*}
\subfigure[]{\includegraphics[scale=0.5]{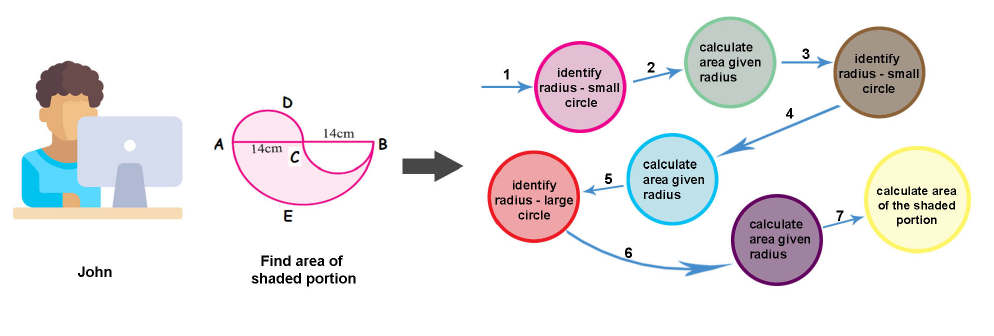}}
\subfigure[]{\includegraphics[scale=0.35]{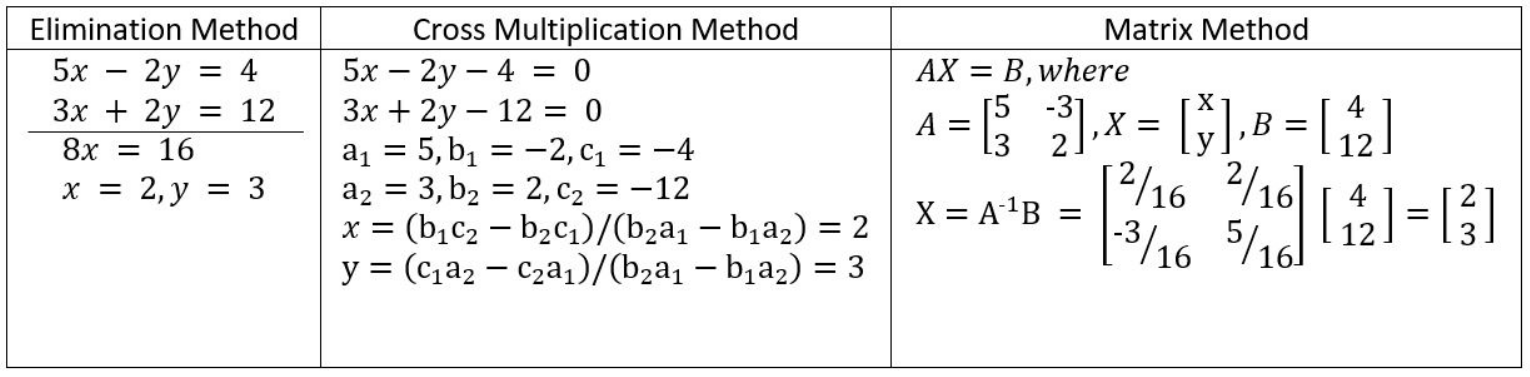}}
\caption{\label{fig:student-strategy} (a) Example of a strategy used by a student to solve a math problem. (b) Different strategies to solve linear equations based on increasing levels of mastery.}
\end{figure*}

The recent pandemic has spurred a remarkable growth in virtual learning and with it, the necessity to develop personalized learning technologies in the absence of face-to-face instruction.
To this end, Intelligent Tutoring Systems (ITSs)~\cite{rus&al13} and more broadly Adaptive Instructional systems (AISs) scale-up one-on-one instruction to large and diverse student populations. 
However, for an AIS to better adapt to a learner, it needs to understand a learner's problem-solving approach.
For example, a student attempting to solve a math problem shown in Fig. \ref{fig:student-strategy} (a) needs to formulate a plan or a {\em strategy} to solve the problem, an example of which is also shown in the figure. By predicting this likely strategy, an AIS can be personalized to meet the student's learning needs such as interventions specific to concepts utilized in the likely strategy.
In this paper, we focus on developing a scalable model to predict Math learning strategies based on interactions between students and educational software. 
In particular, we apply deep neural network based models such as Long-Short Term Memory (LSTMs)~\cite{hochreiter97} to extract sequential patterns that correspond to learner strategies. However, LSTMs are computationally expensive and notoriously slow to converge~\cite{yang&al19} in large datasets. To scale-up, we take advantage of symmetries in strategies. In particular, when we consider the domain of math learning for middle-school students, it is highly unlikely that each student will follow a completely unique strategy. Instead, it is more likely to have groups of students who adopt similar/symmetrical strategies. Further, these symmetry groups are based on the skills or mastery level of a student as illustrated by an example in Fig.~\ref{fig:student-strategy} (b).

To scale-up strategy learning, we develop a {\em non-parametric} approach  to partition the data such that each partition is likely to be invariant in strategies. Thus, training on samples drawn from the partitions effectively represents strategies over the full data. To identify invariance in strategies, we apply DP-Means~\cite{kulisJ12} where we jointly cluster students and problems through the Hierarchical Dirichlet Process (HDP) framework. We adaptively penalize the DP-Means objective such that we have a {\em coarse-to-fine} refinement of the clusters. Specifically, we penalize the formation of clusters based on symmetries induced by the clusters which we estimate through positional embeddings~\cite{vaswani&al17}. To learn a representation for students/problems, we develop a {\em mastery}-based embedding. Specifically, strategies are closely related to mastery over skills, i.e., if a student has mastered a skill, he/she will naturally tend to apply that skill in a strategy. We estimate mastery in a skill based on the attention given to that skill in a Transformer model that maps skills to the correct application of skills.



We evaluate our approach on two datasets from MATHia, a commercial AIS widely used for math learning in schools. The data is available through the PSLC datashop~\cite{StamperKBSLDYS11a}. The $\texttt{Bridge-to-Algebra 2008-09}$ and the $\texttt{Carnegie Learning 2019-20}$ datasets are both large datasets that consist of millions of data instances (each instance is a student-problem pair and has multiple interactions in the dataset). We show that our approach helps scale-up over large datasets while yielding high accuracy in predicting strategies. Further, we show that our {\em mastery}-based embeddings is {\em fair}, i.e., the predictions made by our model is unbiased over diverse levels of mastery.



\section{Proposed Approach}

\eat{
Consider a problem where students solve linear equations. Fig.~\ref{fig:linear-eqn-soln} shows three different {\em strategies} to solve this problem. To adopt the first approach, only a basic understanding of algebra is sufficient, for the second approach, students need a deeper understanding and the final approach while being the most general way to solve the problem also requires more expertise to formulate the problem using linear algebra. Predicting the strategy of a student is of interest to AISs since it allows an intelligent tutor to personalize instruction to a student based on skill levels and intervene appropriately if the student is following a wrong strategy. As seen in the example, the strategy is closely related to the skills/mastery of the student and therefore, we can expect that groups of students with similar mastery are likely to have approximately symmetric strategies. Using this connection, we develop a strategy prediction model by training it with samples over symmetric groups of instances.
}

\subsection{Problem Description}

Since strategy is a generic term, we define it more precisely. Specifically, we consider strategies in the context of structured interaction between students and tutors. In this case, a student interacts with a tutor and solves a problem by sequentially solving the steps that lead to the final solution. Each step is associated with a specific {\em knowledge component} (KC)~\cite{koedinger&al12} which is defined by domain experts and corresponds to the knowledge required to solve that step. Operationally, a step can be associated with multiple KCs in which case, we can just unroll the step to ensure that each step has a single KC. Using this, as in prior work~\cite{ritter&al19}, we define strategies as follows.

\begin{definition}
A strategy used by student $S$ to solve problem $P$ is a sequence of KCs denoted by ${\bf K}$ $=$ $K_1$ $\ldots$ $K_n$, where $K_i$ is a knowledge component that $S$ uses to solve the $i$-th step in $P$.
\end{definition}

Our task is to predict the likely strategy for a given student-problem pair. Note that the same problem can have several possible strategies as shown in Fig.~\ref{fig:student-strategy} (b). Typically, these strategies are closely related to the skill level of a student, i.e., students at higher skill levels may use more complex strategies than students at lower skill levels. We leverage this to learn a representation (embedding) for students and problems such that similar embeddings correspond to symmetries in strategies. We use non-parametric clustering to cluster instances based on this representation and train an LSTM model by sampling from the clusters instead of training over the full dataset. Note that, the use of symmetries to learn better representations from data is in fact a general concept that is formalized under Geometric Deep Learning~\cite{bronstein21}. By training deep models over informative instances from groups of symmetric instances, the Effective Model Complexity is reduced which improves generalization~\cite{nakkiranKBYBS20}.

\subsection{Strategy Invariance}



We define symmetries over ${\bf S}$, ${\bf P}$ which denote the set of students and problems respectively as follows.
\begin{definition}
A {\em strategy-invariant} partitioning w.r.t $\mathcal{D}$ is a partitioning $\{{\bf S}_i\}_{i=1}^{k_1}$ and $\{{\bf P}_j\}_{j=1}^{k_2}$ such that $\forall i,j$, if $S,S'\in{\bf S}_i$, $P,P'\in{\bf P}_j$, $S,S'$ follow equivalent strategies for $P,P'$ respectively.
\end{definition}
where $\mathcal{D}$ is the dataset (sequences of KCs for student-problem pairs), $k_1$ and $k_2$ are the number of partitions/clusters for students and problems respectively.

The benefit of strategy-invariant partitioning is that we can scale up without sacrificing accuracy by training a prediction model only on samples drawn from the partitions instead of the full training data. Therefore, our task is to learn such a partitioning. Since it is hard to know apriori how many partitions are needed, we formulate this as a non-parametric clustering problem and use DP-Means~\cite{kulisJ12} to learn the clusters.

To formalize our approach, we begin with some notation. Let $x_{i1}$ denote the $i$-th student and let $x_{j2}$ denote the  $j$-th problem. Let $z_{ik}$ indicate the student/problem cluster to which $x_{ik}$ belongs. We refer to the student and problem clusters as {\em local} clusters.  A {\em global} cluster combines student and problem clusters. Specifically, $v_{1c}$ $=$ $g$ and $v_{2c'}$ $=$ $g$ indicates that the $c$-th student cluster is associated with the $c'$-th problem cluster through the $g$-th global cluster. The objective function is defined as follows.

\begin{equation}
    \label{eq:dpobj}
    \sum_{p=1}^g\sum_{x_{ij}\in\ell_p}||x_{ij},\mu_p||_2^2+\lambda_{\ell}k+\mathcal{S}(\ell_1\ldots\ell_p)g
\end{equation}

where $\ell_p$ is the $p$-th global cluster, $k$ $=$ $k_1+k_2$ is the total number of local clusters, $g$ is the total number of global clusters, $\lambda_{\ell}$ is a local penalty for the formation of a local cluster and $\mathcal{S}(\ell_1\ldots\ell_p)$ is a global penalty for the formation of a global cluster. Specifically, $\mathcal{S}$ is a measure of symmetry in strategies across the global clusters. To minimize Eq.~\eqref{eq:dpobj},  we can directly apply DP-Means with the Hard Gaussian Hierarchical Dirichlet Process (HDP) clustering as specified in \cite{kulisJ12}. This jointly clusters students and problems while assigning the optimal number of clusters to each. However, in our case, computing the global penalty $\mathcal{S}(\ell_1\ldots\ell_p)$ is challenging, i.e., how can we assign a global penalty that corresponds to the symmetry in the local clusters? We do this using a {\em coarse-to-fine} clustering as follows. We start with a fixed (large) value for the global penalty $\lambda_g$ that results in a coarse clustering. Then, depending upon the symmetries that we estimate from the coarse clusters, we reduce $\lambda_g$ resulting in finer-grained clusters until we reach a fixed point (the change in symmetries is minimal) or for a fixed number of iterations. Algorithm \ref{alg:cfs} summarizes our overall approach for coarse-to-fine clustering. In the next two sections, we describe embeddings (based on mastery) that we learn to represent the data instances in Algorithm \ref{alg:cfs} and the quantification of approximate symmetries in a given clustering.

\begin{algorithm}
\small
    \label{alg:cfs}
    \caption{Coarse-to-Fine Refinement}
    \algsetup{
        linenodelimiter=.
    }
    \begin{algorithmic}[1]
        \renewcommand{\algorithmicrequire}{\textbf{Input:}}
        \renewcommand{\algorithmicensure}{\textbf{Output:}}
        \REQUIRE Student/Problem set: \{$x_{ij}$\}, Penalty parameter: $\lambda_{\ell}$
        \ENSURE  Global strategy clustering \{$\ell_1,\dots,\ell_g$\}\eat{ and number of clusters $g$}
        \\ \textit{Initialize} : Global cluster penalty $\lambda_g = y$ (where $y$ is a large number), $t=0$, cluster coherence $coh_{t-1} = 0$.
        \REPEAT
            \STATE $t = t+1$
            \STATE Perform HDP clustering with parameters: (\{$x_{ij}$\}, $\lambda_{\ell}$, $\lambda_{g}$)
            \STATE Compute cluster coherence score, $coh_t$ = $\mathcal{S}(\ell_1,\ldots\ell_g)$
            \STATE Reduce: $\lambda_{g} = \lambda_{g} - \epsilon$
        \UNTIL{$coh_{t} > coh_{t-1} $}
    \end{algorithmic}
    \label{alg:cfs}
\end{algorithm}

\subsection{Mastery-based Embedding}



Mastery in a skill implies that a student is able to apply the skill correctly  when they have the {\em opportunity} to apply it. For example, if a student has mastered the skill of addition, then the student should be able to apply this in an equation or in a word problem. Mastery has been explored in several well-known learning models, and arguably one of the most well-known methods is Bayesian Knowledge Tracing (BKT)~\cite{corbett01}. Here, mastery over a skill is formalized as a latent variable in an HMM model. This variable is updated based on the performance of the student at each opportunity given to the student to apply that skill. More recently, Deep Knowledge Tracing (DKT)~\cite{piech15} has shown that the hidden layers in a deep network can learn more complex representations of student skills which are much more expressive than BKT. In particular, the connection between symmetries in strategies and mastery is that the ability to use a KC $K$ in a strategy is dependent upon the level of skill in $K'$ that is a pre-requisite to $K$ (also called as pre-requisite structure in DKT). Thus, students with similar pre-requisite structures will tend to use symmetrical strategies. Here, we train an attention model to quantify mastery over a KC and use it to learn an embedding which we refer to as {\tt MVec}.

\begin{figure}
    \centering
    \includegraphics[scale=0.4]{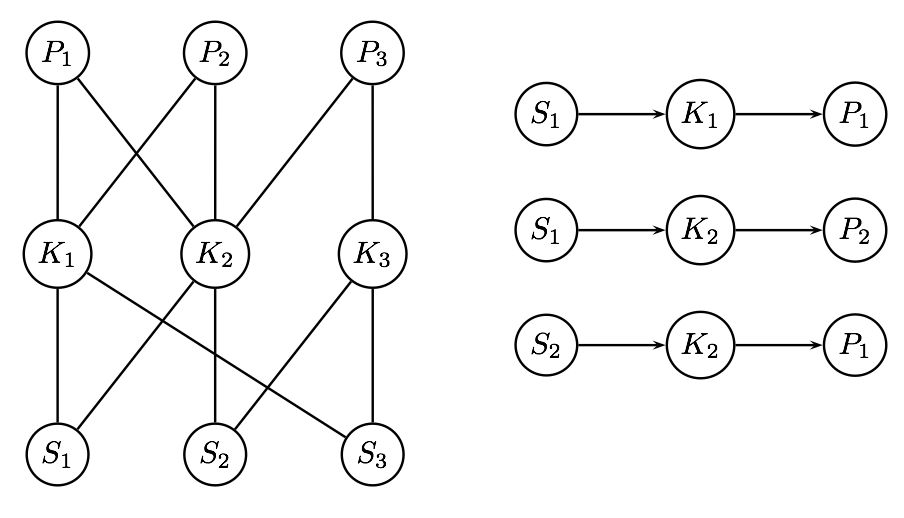}
    \caption{Illustrating a graph network of three students, problems, and KCs. The figure on the right shows some of the sampled random walks/paths.}
    \label{fig:graphex}
\end{figure}

To learn the {\tt MVec} embeddings, we use an approach similar to Node2Vec~\cite{grover&leskovec16}. Specifically, we construct a relational graph $\mathcal{G}$ $=$ ({\bf V},{\bf E}) as follows. Each student, problem, and KC in the training data is represented as a node $V\in{\bf V}$. For every student $S$ who uses KC $K$ as a step to solve problem $P$, there exists 2 edges $E, E'$ $\in$ {\bf E}, where $E$ connects the node representing the student to the node representing the KC and $E'$ connects the node representing the KC to the node representing the problem. An example graph over 3 students, problems, and KCs is shown in Fig.~\ref{fig:graphex}. We now sample paths in the graph similar to Node2Vec and learn embeddings for these paths using word embedding models (Word2Vec)~\cite{mikolov&al13}. Specifically, the objective function is as follows.

\begin{equation}
    \max_{f}\sum_{u\in {\bf V}}log P(N_Q(u)|f(u))
\end{equation}

where $f:u\rightarrow \mathbb{R}^d$ is the representation for nodes $u\in{\bf V}$, $N_Q(u)$ denotes the neighbors of $u$ sampled from a distribution $Q$. As in Node2Vec, we assume a factorized model that gives us the conditional likelihood that is identical to Word2Vec.

\begin{equation}
    P(n_i|f(u))=\frac{\exp(f(n_i)\cdot f(u))}{\sum_{v\in{\bf V}}\exp(f(v)\cdot f(u))}
\end{equation}

where $n_i$ is a neighbor of $u$. The conditional likelihood is optimized by predicting neighbors of $u$ using $u$ as input in an autoencoder neural network. The hidden layer learns similar embeddings for nodes with symmetrical neighborhoods. To do this, we generate walks on $\mathcal{G}$ as shown for the example in Fig.~\ref{fig:graphex}, and in each walk, given a node, we predict neighboring nodes similar to predicting neighboring words in sentences. To generate these walks, a simple sampling strategy $Q$ is to randomly sample a neighbor for a node. However, in our case, it turns out that each neighbor may have different importance when it comes to determining symmetry. Specifically, if a student has achieved mastery in applying a KC to a problem, then the corresponding edges should be given greater importance when determining symmetry between nodes in $\mathcal{G}$. To do this, we train a Sequence-to-Sequence attention model~\cite{vaswani&al17} from which we estimate the sampling probabilities for edges in $\mathcal{G}$.

\begin{figure*}
    \centering
    \scalebox{1.0}{
        \includegraphics{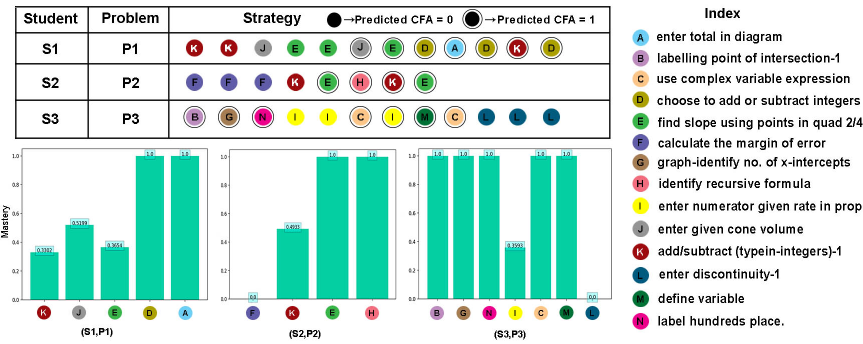} 
    }
    \caption{An example to illustrate the use of attention for mastery estimation. The bar charts show for each KC, the attention on a KC across steps that the student solves successfully (CFA=1) normalized by total attention for that KC. Larger values indicate that the model believes the student understands the KC as the attentions on it is large when CFA=1 and vice versa.}
    \label{fig:latent-mastery}
\end{figure*}

The intuitive idea in quantifying mastery is illustrated in Fig.~\ref{fig:latent-mastery} which shows the opportunities given to 3 students to apply KCs in different problems. For each sequence of KCs, we predict if the student got the step correct or wrong in the first attempt (abbreviated as CFA for Correct First Attempt) when given an opportunity to apply the KC. The CFA values are performance indicators for the student, i.e., if they have mastered a KC, then they are likely to get the step correct in every opportunity they get to apply that KC. We train a model to predict the CFA values (CFA $=$ 1 indicates a correct application of the KC) given the KCs used in a problem. The predicted values from the model are shown for each KC. The bar graphs show mastery over the KCs. As seen here, the first student is inconsistent in applying the skill, {\em find the slope using points} (labeled as $E$) since the predictions for this oscillate between 0 and 1 whenever the student tries to apply this KC. On the other hand, student 2 consistently applies the same skill correctly and therefore the attention value is higher. We train the attention model from opportunities based on curriculum structure. Specifically, the curriculum consists of multiple units and each unit is further subdivided into sections.
For each student $S$, from every unit that the student has completed say $U$, we select a problem $P$ from each section that the student has worked on in $U$ and train the model to predict the CFA values for each KC used in $P$. 
We use the standard architecture described in ~\cite{vaswani&al17} for this model. Specifically, the input consists of the KC sequence, and the encoder maps this sequence to a latent representation and the decoder decodes the CFA values one at a time. The attention is given by,
\begin{equation}
\label{eq:atteq}
    Attention(Q,Y,V) = softmax\left(\frac{QY^T}{\sqrt{d_k}}\right)V
\end{equation}
where $Q$, $Y$, and $V$ are the standard query, key, and value matrices respectively as defined in~\cite{vaswani&al17}, and $d_k$ is the dimension of the embedding that represents the latent representations. We use the encoder-decoder attention, i.e., the query is the decoder representation and the key is the encoder representation. The attention weights are an estimate of the alignment between encoded latent representations of mastery with the decoded representation of correctly applying a skill at each step in the problem. The projection of mastery over $K$ based on the attention vectors is estimated by the following equation.



\begin{equation}
\label{eq:attn}
  \alpha(S,P,K) = \frac{\sum_i\sum_{v\in\pi(a_i)}v}{\sum_i\sum_{v\in\pi(a_i)}v+\sum_i\sum_{v\in\bar{\pi}(a_i)}v}  
\end{equation}

where $\pi(\cdot)$ extracts only those values in the input vector where the corresponding output for that step is predicted as 1, i.e., the model predicted that the student could solve the step correctly. $\bar{\pi}(\cdot)$ is the complement of $\pi(\cdot)$, i.e., it extracts attention values corresponding to steps that were predicted as mistakes. We sample $\mathcal{G}$ as a factored distribution, i.e., $Q(S)*Q(K|S)*Q(P|K,S)$, where $Q(S)$ is the probability of sampling a student node, $Q(K|S)$ is the probability of sampling a KC $K$ given student $S$ and $Q(P|K,S)$ is the probability of sampling problem $P$ given $K,S$. We assume that $Q(S)$ is a uniform distribution over students. The conditional distributions are defined as follows.
\begin{equation}
\label{eq:q1}
Q(K|S)=1/n\sum_p\alpha(S,P,K)    
\end{equation}
\begin{equation}
\label{eq:q2}
Q(P|K,S)=\alpha(S,P,K)
\end{equation}
\eat{
$$
Q(x|y) = \begin{cases}
  1/n\sum_p\alpha(y,p,x)  &  \text{ y is a student node} \\
  1/m\sum_u\alpha(u,x,y) & \text{y is a KC node}
\end{cases}
$$
}
where $n$ is the number of opportunities given to student $S$ to apply KC $K$. The algorithm to generate {\tt MVec} embeddings is shown in Algorithm \ref{alg:mvec}. As shown here, we sample a path in the graph as follows. We first sample student $S$ uniformly at random, then we sample a KC $K$ from $Q(K|S)$ and a problem from $Q(P|K,S)$. We then predict each node in the path using the neighboring nodes through a standard Word2Vec model. The resulting embeddings are learned in the hidden layer of this model. Note that for scalability, we do not construct/store the full graph $\mathcal{G}$ at any point. Instead, we only sample paths in an online manner as shown in Algorithm \ref{alg:mvec}.


\begin{algorithm}
\small
    \caption{Generate MVec embeddings}
    \algsetup{
        linenodelimiter=.
    }
    \begin{algorithmic}[1]
        \renewcommand{\algorithmicrequire}{\textbf{Input:}}
        \renewcommand{\algorithmicensure}{\textbf{Output:}}
        \REQUIRE Relation Graph: $\mathcal{G} = (\bf V, \bf E)$ with student, problem and KCs as nodes, Embedding dimension: $d$, pre-trained attention-model $\mathcal{A}$
        \ENSURE Embeddings for each node $v \in \mathbb{R}^d$
        \\ \textit{Initialize}: set of walks, $\mathcal{W} = empty$
        \FORALL{$t$ $=$ 1 to $T$}
        \STATE Sample  a path $<S,K,P>$ in $\mathcal{G}$ from $Q(S)*Q(K|S)*Q(P|K,S)$ using Eq.~\eqref{eq:q1} and \eqref{eq:q2}.
        \STATE $\mathcal{W}$ $=$ $\mathcal{W}$ $\cup$ $<S,K,P>$
        \ENDFOR
        \STATE$v_e = word2vec(\mathcal{W}, d)$
        \RETURN $v_e$
    \end{algorithmic}
    \label{alg:mvec}
\end{algorithm}


\begin{table*}[]
    \centering
    \caption{Illustrating symmetries in strategies where similar colors indicate similar steps. The three strategies have a different number of steps and other small differences but the core idea behind them remains the same. A naive matching will however assume that the three strategies are very different.}
    \resizebox{1\textwidth}{!}{
    \begin{tabular}{|c|c|c|c|c|c|c|c|c|}
        \hline
        \cellcolor{gray!25}\textbf{Strategy 1} & action add/sub & \cellcolor{red!15}params add/sub & \cellcolor{blue!10}rm-extra-terms & \cellcolor{red!30}rm-coeff & act mul/div & \cellcolor{red!55}params mul/div & \cellcolor{yellow!30}rm-extra-terms & \cellcolor{orange!80}find-x\\
        \hline
        \cellcolor{gray!25}\textbf{Strategy 2} & & & \cellcolor{blue!25}rm-constants & \cellcolor{red!30}rm-coeff & & \cellcolor{red!50}div-coeff-of-x & \cellcolor{yellow!30}rm-extra-terms & \cellcolor{orange!80}find-x\\
        \hline
        \cellcolor{gray!25}\textbf{Strategy 3} & & \cellcolor{red!30}rm-coeff & & \cellcolor{red!50}div-coeff-of-x & simplify-eqn & rm-constants & \cellcolor{yellow!30}rm-extra-terms & \cellcolor{orange!80}find-x\\
        \hline
    \end{tabular}
    }
    \label{tab:position_similarity}
\end{table*}

\subsection{Strategy Symmetry}
\label{section:strategy_symmetry}

To perform the coarse-to-fine refinement (Algorithm 1), we need to estimate symmetry in strategies across the global clusters, i.e., $\mathcal{S}(\ell_1,\ldots,\ell_g)$. A naive approach is to represent the sequence of KCs as a single vector by taking a mean of their {\tt MVec} embeddings. In this case, we can estimate the symmetry between the two strategies based on the distance of their mean vectors. However, this representation assumes that all permutations of a sequence of KCs are equivalent to each other. This is clearly problematic since an essential aspect of reasoning about problems is to understand the sequence of steps. On the other hand, if we strictly require the KCs from two strategies to align exactly, this is too restrictive since it is perfectly acceptable for students to add/delete steps in reasoning though the core strategy remains equivalent. An example of this is illustrated in Table~\ref{tab:position_similarity}. In this case, 3 strategies are shown which differ in their steps. However, the core idea behind each of the strategies is the same and it turns out that aligning the key steps makes the strategy invariant across the students as shown in the figure.


To match strategies approximately, we represent a strategy using a combination of embeddings and positional encodings~\cite{vaswani&al17} and approximately align two strategies to estimate the symmetry between them. The idea in positional encodings is to encode positions in a sequence using a continuous vector space with sine and cosine functions. Given the positional encoding vectors for say the $i$-th KC in strategy $S_1$ and  the $j$-th KC in strategy $S_2$, their similarity $\vec{p}_i^\top \vec{p}_j$ will indicate how close or far apart these positions are. Note that this similarity takes into account the variability in the number of steps in $S_1$ and $S_2$. Thus, the step-positions in the two strategies will have similar vector representation if they are at similar positions relative to their overall sequences.

A KC $K$ in the strategy is represented by its {\em positional embedding} $\vec{K}$ $=$ $\vec{K_e}$ $+$ $\vec{K_p}$ where $\vec{K_e}$ is the embedding for $K$ and $\vec{K_p}$ is the positional encoding for $K$ in the strategy. To compute symmetry between strategies, we compute an alignment between their positional embeddings.
Alignment is a fundamental problem in domains such as Bioinformatics where a classical approach that is often used is the Smith and Waterman algorithm (SW)~\cite{smith_waterman}. The idea is to perform local search to compute the best possible alignment between two sequences. SW requires a {\em similarity function} which in our case is the similarity between two KCs and we set this to be $s(K,K')$ $=$ $\vec{K}^\top\vec{K}'$, i.e., the cosine similarity between the embeddings of $K$ and $K'$. Further, SW also requires a {\em gap penalty} which refers to the cost of leaving a gap in the alignment. In our case, we set the gap penalty to 0 since we want symmetry between strategies to be invariant to gaps. That is, if two strategies are symmetric, adding extra steps in the strategies is acceptable. SW iteratively computes a scoring matrix based on local alignments. The worst-case complexity to compute the scoring matrix that gives us scores for the best alignment is equal to $O(m*n)$ where $m$ and $n$ are lengths of the strategies. Note that in our case, we are interested in quantifying the symmetry based on the alignment. Specifically, let ${\bf K}$ and ${\bf K}'$ be two strategies of lengths $n$ and $m$ respectively. SW gives us an alignment between ${\bf K}$ and ${\bf K}'$ denoted by $L({\bf K},{\bf K}')$. The alignment can be represented by pairs of aligned KCs in the strategy or a gap. We compute the symmetry score between ${\bf K}$ and ${\bf K}'$ as $r({\bf K},{\bf K}')$ $=$ $\frac{1}{\max(n,m)}$ $\sum_{(K,K')\in L({\bf K},{\bf K}')}$ $(\vec{K}^\top\vec{K'})$, where $(K,K')$ $\in$ $L({\bf K},{\bf K}')$ are aligned KCs and $\vec{K}^\top\vec{K'}$ is their similarity. We see that $0\leq$ $r({\bf K},{\bf K}')$ $\leq 1$. Based on this, we estimate symmetry in the clustering as follows.
\eat{
\begin{align}
\label{eq:sym}
    \mathcal{S}(\ell_1,\ldots,\ell_g)=
    &\frac{1}{g}\sum_{p=1}^g\frac{2}{|T(\ell_p)| (|T(\ell_p)|-1)}\\\nonumber
    &\sum_{{\bf K},{\bf K'}\in T(\ell_p)}r({\bf K},{\bf K}')
\end{align}
}
\begin{align}
\label{eq:sym}
    \mathcal{S}(\ell_1,\ldots,\ell_g)=
    &\frac{1}{g}\sum_{p=1}^g \mathbb{Z}_p \sum_{{\bf K},{\bf K'}\in T(\ell_p)}r({\bf K},{\bf K}')
\end{align}
$T(\ell_p)$ is a set of all strategies in $\ell_p$, i.e., $\forall S,P\in\ell_p$, where $S$ is a student and $P$ is a problem, $\exists {\bf K}\in T(\ell_p)$, where ${\bf K}$ is the strategy followed by $S$ for $P$ and $\mathbb{Z}_p=\frac{2}{|T(\ell_p)|(|T(\ell_p)|-1)}$ is the normalization term. Thus, a larger value of $\mathcal{S}(\ell_1,\ldots,\ell_g)$ implies that the clustering corresponding to $\ell_1,\ldots,\ell_g$ has a greater degree of symmetry in strategies.

\subsection{Training the Prediction Model}
We use the LSTM architecture in ~\cite{shakya2021} to predict strategies. Specifically, the model is a one-to-many LSTM that takes the embeddings corresponding to a student and a problem as input and generates a sequence of knowledge components as output. To train the model, we sample instances from the converged global clusters, and for each sampled student-problem pair, we concatenate their positional embeddings and feed them as input to the LSTM. The output corresponds to a sequence of KCs, each of which is encoded as a one-hot vector. To handle variable-length strategies, a special {\em stop} symbol is used to denote the end of a sequence. The entire model is trained using the categorical cross-entropy loss.

\section{Experiments}

\begin{table}[]
    \centering
    \caption{Main parameters for the models.}
    \label{tab:params}
    \resizebox{0.48\textwidth}{!}{
    \begin{tabular}{c||c}
        \hline
        {\bf Transformer-based model} & {\bf LSTM-based strategy model}\\
        \hline\hline
        Dimension $\rightarrow$ 512 & Latent Dimension $\rightarrow$ 200\\
        Number of layers $\rightarrow$ 6 & Epochs $\rightarrow$ 60\\
        Number of heads $\rightarrow$ 8 & Batch Size $\rightarrow$ 30\\
        Dimensions of key, value and query $\rightarrow$ 64 & Adam Optimizer with Learning rate 0.01\\
        Max Sequence Length $\rightarrow$ 150  & Dropout $\rightarrow$ 0.1\\
        Dropout $\rightarrow$ 0.1 &\\
        Weight Sharing $\rightarrow$ False &\\
        \hline
    \end{tabular}
    }
\end{table}

\begin{figure*}
    \centering
    \subfigure[]{\scalebox{0.25}{
        \includegraphics{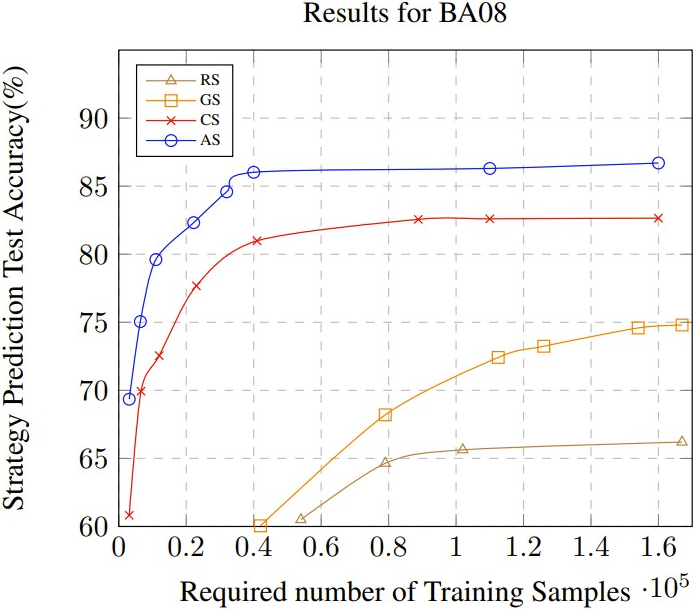}
    }}
    \subfigure[]{\scalebox{0.25}{
        \includegraphics{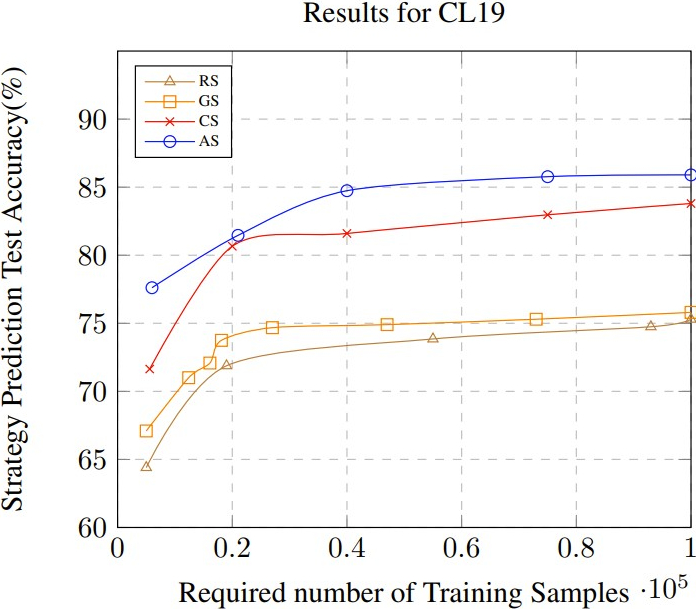}
    }}
    \subfigure[]{\scalebox{0.25}{
        \includegraphics{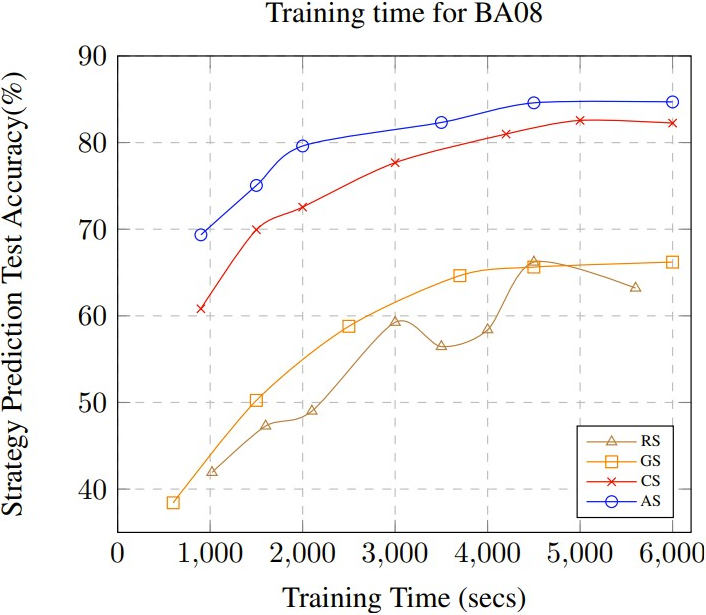}
    }}
    \\
    \subfigure[]{\scalebox{0.25}{
        \includegraphics{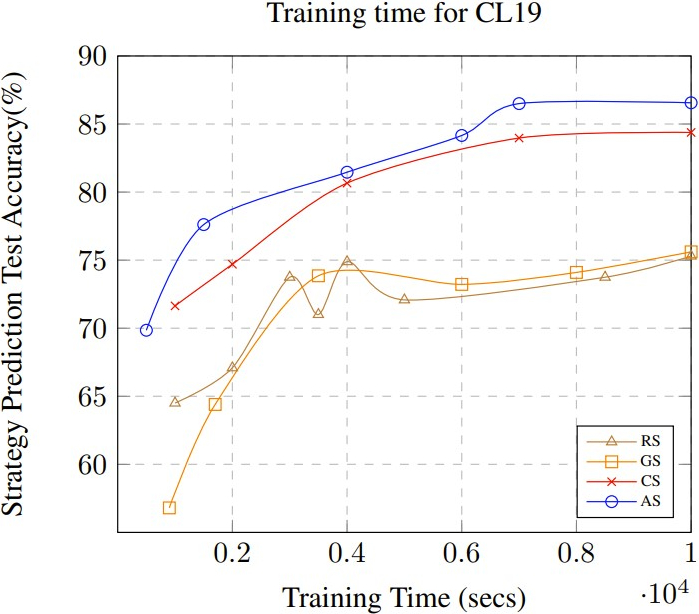}
    }}
    \quad
    \subfigure[]{\scalebox{0.21}{
        \includegraphics{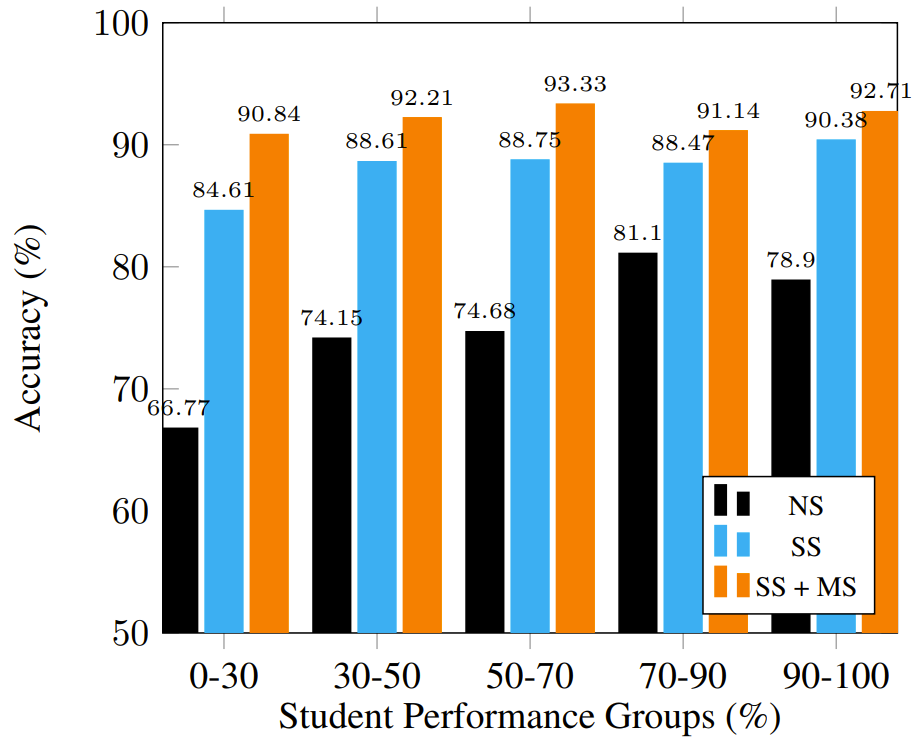}
    }}
    \caption{Illustrating Scalability vs Accuracy. (a), (b) show test accuracy for strategy prediction for varying training datasize limits.  (c), (d) show accuracy (strategy prediction) for different training time limits. (e) shows accuracy for different groups of students (based on their performance). The x-axis denotes different ranges of \%s, where a range $a-b$ denotes that students in this group got $>a$ and $<b$ steps correct in their first attempt. The y-axis shows accuracy over the groups.}
    \label{fig:stratpred}
\end{figure*}

\subsection{Experimental Setup}
We used two datasets provided by MATHia for evaluating our proposed approach, Bridge-to-Algebra 2008-09 (\texttt{BA08}) and Carnegie Learning MATHia 2019-20 (\texttt{CL19}). \texttt{BA08} consists of about $20$ million interactions for about $6000$ students and $52k$ unique algebra problems and $1.6$ million \textit{instances} (student, problem pairs). \texttt{CL19} contains $47$ million interactions for $5000$ students, $32k$ problems and $1.9$ million \textit{instances}. Both datasets are publicly available through the PSLC datashop.
To train the attention model, we used the implementation in~\cite{vaswani&al17}. For the strategy prediction, we used the LSTM model as described in the previous section. The parameters for the two models are shown in Table \ref{tab:params}. For generating {\tt MVec} embeddings, we used Gensim~\cite{mikolov&al13} with an embedding dimension set to 300. We initialize the local cluster penalty $\lambda_{\ell} = 7$ and initial global cluster penalty $\lambda_g = 9$ for Coarse-to-Fine refinement. We perform our experiments with 64 GB RAM and a 16GB quadro GPU.

\subsection{Comparision to Baselines}
We compared our approach with the following methods. The first one is the approach in~\cite{shakya2021} ($\mathrm{CS}$) where the LSTM is trained over samples, but mastery is not used to learn the embeddings. We also applied a general importance sampling approach for deep learning proposed in~\cite{KatharopoulosF19} ($\mathrm{IS}$) using their publicly available implementation. However, $\mathrm{IS}$ failed to output any results for datasets of our size and therefore we do not show it in our graphs. We also developed a stratified sampler ($\mathrm{GS}$ for group sampling) where the distribution is only proportional to the number of problems solved by a student, i.e., we sample more instances from students that have data associated with them. The last baseline is a naive Random Sampler ($\mathrm{RS}$) used as a validation check where we sample students and problems uniformly at random. We refer to our approach as Attention Sampling ($\mathrm{AS}$).
In our evaluation, for each approach, we enforce a limit on the number of training instances and measure the test accuracy of models trained with this limit. This is similar to {\em effective model complexity}~\cite{nakkiranKBYBS20} which is the number of training samples to achieve close to zero error. We report the average accuracy based on three training runs.

\eat{

\begin{table}[]
    \centering
    \caption{Ablation study with NS (No symmetries used), SS (Symmetries without using mastery) and MS (Adding the mastery model to better identify symmetries). Results are shown for 2 datasets with different sample sizes. Accuracy results in \%.}
    \begin{tabular}{|c|c|c|c|c|c|c|}
        \hline
         \multirow{2}{*}{\makecell{
         \backslashbox[24mm]{Expts.}{\makecell{Dataset}}}} & \multicolumn{3}{c|}{\textbf{BA08}} & \multicolumn{3}{c|}{\textbf{CL19}}  \\
         \cline{2-7}
         & \textbf{40k} & \textbf{100k} & \textbf{150k} & \textbf{40k} & \textbf{80k} & \textbf{100k} \\
         \hline
         \textbf{NS} & 60.05 & 71.14 & 74.58 & 74.81 & 75.4 & 75.8\\
         \hline
         \textbf{SS} & 80.98 & 82.3 & 82.65 & 81.6 & 83.2 & 83.8\\
         \hline
         \textbf{SS + MS} & {\color{Green} 86.02} & {\color{Green} 86.21} & {\color{Green} 86.53} & {\color{Green} 84.74} & {\color{Green} 85.8} & {\color{Green} 85.9}\\
         \hline
    \end{tabular}
    
    \label{tab:ablation}
\end{table}
}

\subsection{Results}
\noindent{\bf Accuracy and Scalability.} The accuracy results for BA08 and CL19 are shown in Fig.~\ref{fig:stratpred} (a) and (b). As shown in (a), for BA08, in our approach ($\mathrm{AS}$), it takes less than around $10\%$ of the data to obtain test accuracy that is greater than 80$\%$. $\mathrm{CS}$ is the next best performer but is consistently below $\mathrm{AS}$ for all training sizes. $\mathrm{GS}$ performs significantly worse which illustrates that symmetries are more complex and a simple grouping based on problems/students is insufficient. The poor performance of $\mathrm{RS}$ validates that the problem of choosing the correct samples is indeed a challenging one. As seen in Fig.~\ref{fig:stratpred} (b), for a considerably larger dataset CL19, we can observe similar performance as in BA08. $\mathrm{AS}$ remains the best performer and here $\mathrm{CS}$ is less stable since we see a performance drop as we increase the limit on training samples. This suggests that $\mathrm{CS}$ may not be able to capture all symmetries (at all mastery levels) and thus may produce a less effective training sample. The results for $\mathrm{GS}$ and $\mathrm{RS}$ are similar to those observed in BA08. As mentioned before, $\mathrm{IS}$ failed to produce any results. Fig.~\ref{fig:stratpred} (c) and (d) show the training time required to obtain a specific accuracy for BA08 and CL19 respectively. Even with the extra processing that is needed to compute the mastery-based embeddings and the non-parametric clustering, $\mathrm{AS}$ requires the shortest training time to achieve an accuracy that is higher than the other approaches. This illustrates the significance of leveraging symmetries in the data to train the model. As mentioned before, the full data is infeasible to train and when attempting to use the full data, the model did not converge even after several days of training time. For CL19, the training time is larger since it takes longer to compute the groups using non-parametric clustering due to the larger size of the dataset. However, considering that CL19 is significantly larger than BA08, we see that $\mathrm{AS}$ can easily scale big datasets.

\noindent{\bf Analyzing Bias.} The strategy prediction model is fair if it does not disproportionately favor a sub-group of students, i.e., if it strategy predictions are not biased for students at a specific mastery-level. To evaluate fairness, we divided the students in the test data into 5 performance groups. The performance groups are based on the $\%$ of problem steps the students solve correctly in their first attempt. For a student $S$ in performance group ${\bf G}$, we predict the strategies for all problems attempted by $S$ in the test set and measure the average accuracy $\mu_S$. We then compute the accuracy over a performance group as $1/|{\bf G}|\sum_{S\in{\bf G}}\mu_S$. Fig.~\ref{fig:stratpred}(e) shows our results for variants, NS (no symmetries are used), SS (approach in~\cite{shakya2021} that does not use mastery) and SS+MS (our approach that uses mastery). As shown here, SS+MS yields a consistent improvement in performance across all performance groups illustrating that our approach is fair across diverse mastery levels.
\section{Conclusion}
Mastery over concepts is closely related to problem-solving strategy. Using this connection, here, we developed embeddings and through non-parametric clustering on these embeddings, we segregated big datasets into partitions where strategies are likely to be invariant. Sampling from these partitions scales up strategy prediction such that it performs equally well for students at all skill levels. We demonstrated the effectiveness of our approach in Math learning datasets. 

\bibliography{main}
\end{document}